\documentclass[a4paper,12pt]{article}
\usepackage[backend=bibtex,bibstyle=ieee,sorting=none,citestyle=numeric-comp]{biblatex}
\usepackage{pgf,tikz,pgfplots,tkz-euclide,tikz-cd}
\usetikzlibrary{arrows,positioning,calc,fadings,decorations.pathreplacing,decorations}
\usepackage{amssymb,amsmath,amsthm,mathtools}
\usepackage{subcaption,caption}
\usepackage{hyperref}
\usepackage{enumerate}
\hypersetup{pdfpagemode={UseOutlines},
bookmarksopen=true,
bookmarksopenlevel=0,
hypertexnames=false,
colorlinks=true,
citecolor=black,
linkcolor=black,
linkbordercolor={1 0 0},
urlcolor=black,
pdfstartview={FitV},
unicode,
breaklinks=true,
}
\usepackage[affil-it]{authblk}
\usepackage[title]{appendix}
\newcommand\comment[1]{#1}
\def\ba{\mathbf{a}}
\def\R{\mathbf{R}}
\def\Z{\mathbf{Z}}
\def\C{\mathbf{C}}
\def\H{\mathbf{H}}

\def\X{\mathcal{X}}
\def\M{\mathcal{M}}
\def\bx{\mathbf{x}}
\def\L{\mathcal{L}}
\def\sA{A_{\sigma}}
\def\bA{A_{\Bar{\sigma}}}
\def\bsigma{\Bar{\sigma}}

\def\bell{\ell}
\newcommand\G[1]{\Gamma\left(#1\right)}

\newcommand\eq[1]{(\ref{#1})}
\def\pa{\partial}
\def\map{\longmapsto}

\def\conf{{\mathcal{C}}}
\def\m{{\boldsymbol\mu}}
\def\bxi{{\boldsymbol\xi}}

\newcommand\inmu[3]{I^{{#1}}_{{#2}}({#3})}

\newcommand\rt{\longrightarrow}

\theoremstyle{definition}

\def\conf{\mathcal{C}}
\def\m{\boldsymbol{\mu}}

\newcommand\cl[1]{Cl_{#1}(\R)}
\newcommand\cls[1]{Cl^{\star}_{#1}(\R)}
\def\e{e}

\def\Ga{\Gamma}

\def\bxi{\boldsymbol{\xi}}

\def\mA{\mathcal{A}}
\renewcommand{\bar}{\overline}

\newcommand\cone[1]{\operatorname{cone}(#1)}
\def\appell(#1,#2,#3,#4,#5,#6){F_4\big(#1,#2,#3,#4;\ #5,#6\big)}
\newcommand{\tr}{\operatorname{Tr}}
\title{Conformal integrals in all dimensions as generalised hypergeometric 
functions and Clifford groups}
\author{Aritra Pal\thanks{intap@iacs.res.in}}
\author{Koushik Ray\thanks{koushik@iacs.res.in}}
\affil{Indian Association for the Cultivation of Science,\\
Kolkata - 700032, India.}
\begin{document}
%
%
%
%
%
\maketitle
\thispagestyle{empty}
\begin{abstract}
\noindent
Euclidean conformal integrals for an 
arbitrary number of points in any dimension are evaluated. 
Conformal transformations in the Euclidean space 
can be formulated as the 
M\"obius group in terms of Clifford algebras. 
This is used to interpret conformal integrals as
functions on the configuration space of points 
on the Euclidean space, solving linear differential equations,
which, in turn, is related to toric GKZ (Gelfand-Kapranov-Zelevinsky) systems.
Explicit series solutions for the conformal integrals are obtained 
using toric methods as GKZ hypergeometric functions. 
The solutions are made symmetric under the action of permutation of 
the points, as
expected of quantities on the configuration space of unordered
points, using the monodromy-invariant unique Hermitian form.  
Consistency of the solutions among different number of points is shown. 
\end{abstract}
\clearpage
\section{Introduction}
\noindent In this article we study conformal integrals \cite{Sym}
which appear, among others instances, in 
the computation of conformal correlation functions. 
Conformal integrals appearing in the conformal
block expansions
have been studied extensively using different techniques
\cite{DO1,DO2,DO3,CPP,HR,KKS,EIK,FMP,Ros,Par1,Par2,For1,For2,Par3,For3,
For4,SOM1,SOM2,SOM3,SOM4,For5}. 
They furnish 
representations of the conformal group \cite{PR1,PR2}. 
An $N$-point conformal integral in the $D$-dimensional Euclidean space
$\R^D$ is an integral over the whole space involving $N$ marked points 
transforming under the global conformal group $SO(1,D+1)$.
While the two-point and three-point 
integrals are completely fixed by the conformal group, evaluation of
four and higher point integrals are
rendered difficult by the occurrence of conformally invariant quantities,
namely, the cross ratios. 
Let us recall that if a Lie group 
$G$ acts transitively on a manifold $M$ 
as $G\times{M}\longrightarrow{M}$, $(g,x)\map g(x)$, for each $g\in G$ and 
each $x\in{M}$, then a representation $T$ of 
the group is defined by lifting the geometric action of $G$ to a   
class of ``regular" functions $f$ on ${M}$  
by $T_gf(x)=f(g^{-1}x)$. An $N$-point 
conformal integral is a representation of the conformal group 
in this sense, where ${M}$ 
is the configuration space of $N$ points on $\R^D$. The group $G$ is the  double cover of the rotation group $SO(1,D+1)$, which from 
now on will be referred to as the 
conformal group. According to Liouville's theorem, 
the geometric action of the conformal group on the 
configuration space is given by a M\"obius transformation. In two dimensions
the conformal group is $SL(2,\C)$,  which induces the M\"obius transformation
on each point through a fractional linear transformation. In four
dimensions the conformal group is $SL(2,\H)$, where $\H$ denotes the space
of quaternions \cite{porteous,wilker}. Generalizing, in $D$ dimensions the 
conformal group is written in terms of a subgroup of invertible elements of
the Clifford algebra of $\R^D$, namely, $\cl{D}$ \cite{Ahlf,loun}. 
In here, we concern ourselves with
this description of the conformal group as the M\"obius group, having 
a geometric action on $\R^D$. Amongst the various models of the configuration
space of points we adopt the Fulton-Macpherson completion of the configuration 
space of $N$ unordered non-coalescing points in $\R^D$ \cite{FM}. 
As for the regular functions, we look for the
representation of the M\"obius group among the functions on this model 
of the configuration space.
In two dimensions, the configuration space pertains to that on the complex 
projective space. In this case, conformal integrals, along with the
pairwise differences of the coordinates of the $N$ points furnish 
a linear system on the configuration space, obtained as solutions to
a Lauricella system \cite{looi}. Multi-valuedness of the integrals 
is incorporated by describing it through the germs of solutions of the 
Lauricella differential equations. A generalization of this has been 
obtained earlier for the 
four-dimensional case. A set of differential equations generalizing 
the Lauricella system is deduced by expressing the conformal integrals
in terms of quaternions, thereby making their covariance under $SL(2,\H)$
manifest \cite{PR1}. The conformal integrals are 
translation-invariant homogeneous solutions of
a set of differential equations. These are evaluated by exhibiting that 
the differential equations
are solved by the
solutions of a set of GKZ (Gelfand-Kapranov-Zelevinsky) 
hypergeometric equations \cite{PR2}. 
In this approach, it is the Lauricella-like differential equation that 
fixes the dependence on the cross ratios. 

In here, this approach is generalized to $\R^D$ for any $D$, using 
the formulation of the conformal group as the M\"obius group in terms
of a Clifford algebra \cite{Ahlf,loun}. 
We write  the conformal integral in terms
of certain elements of the Clifford algebra, called the Clifford numbers.
We derive a set of differential equations for the conformal integral 
in terms of the Clifford numbers. The equations are of the same
form as the ones obtained for quaternions, with the latter replaced 
with Clifford numbers. Solutions of the set are then given by 
the GKZ hypergeometric functions obtained earlier \cite{PR2}, 
upon appropriate amendments in the dimensionality, from $4$ to $D$ 
in the formul\ae. 

The GKZ system furnishes a local system whose solutions are the germs of 
the conformal integral. Solutions are obtained
as series with domains of convergence restricted by the
principal component of the discriminant locus of the GKZ system, in the
space of the cross ratios. 
While the conformal integral, being the sheaf of germs of the
GKZ hypergeometric functions, is per se symmetric under the 
permutation of the $N$ points, the process of solving  differential equations
breaks the permutation 
symmetry partially, through the choice of cross ratios. 
The permutation symmetry is restored by adding the series, treated as a basis,
with appropriate constants, through analytic continuation.
The Fulton-Macpherson configuration space of unordered points possesses an
action of the group of permutations of the points. This translates to 
analytic continuations of solutions, related to each other by monodromy, which,
in turn, is a representation of the fundamental group of the configuration
space. The fully symmetric solution, then, is the unique monodromy-invariant 
Hermitian form \cite{beu,mheo2,ver}.

In section~\ref{mobtrans} we define the M\"obius group in the parlance 
of Clifford algebra \cite{Ahlf,loun}. In section~\ref{repmob} we then 
express the conformal 
integral \cite{Sym} in the corresponding notation in \eq{inmu} and relate it to
the configuration space of points. The set of differential equations 
satisfied by the conformal integral are obtained and expressed in terms
of a GKZ hypergeometric system. These are solved to obtain 
expressions for the conformal integral in section~\ref{ex456}.
We close with discussions on several 
aspects of the present formulation. 

%
\section{M\"obius Transformations and Clifford group}
\label{mobtrans}
Let us begin with a discussion of some aspects of Clifford algebras relevant for the 
M\"obius group. The Clifford algebra $\cl{D}$ is a real associative 
algebra generated by 
\begin{equation}
\label{cliff}
\{e_{\nu}|e_{\nu}e_{\nu'}+e_{\nu'}e_{\nu}=-2\delta_{\nu\nu'},\,
\nu,\nu'=1,2,\cdots,D\}.
\end{equation}
It is a vector space of dimension $2^D$. A basis for it is given by 
$e_0=1$ and the monomials, also called blades, formed from the generators, such as
\begin{equation}
\e_{\nu_1}\e_{\nu_2}\cdots\e_{\nu_p},  
\end{equation}
where $1\leqslant p\leqslant D$ and the indices  
are arranged in a strictly
increasing order.  The degree of the 
monomial furnishes the grade of the elements of the basis, $p$ for the above 
element, for example.
An element of $\cl{D}$ is expressed as a linear combination of the $2^D$ 
basis elements as
\begin{equation}
\label{xpand}
\X = x^0e_0+
\smashoperator{\sum_{\substack{p=1\\1\leq \nu_1<\nu_2<\cdots<\nu_p
\leq D}}^D}  x^{\nu_1\nu_2\cdots\nu_p} 
\e_{\nu_1}\e_{\nu_2}\cdots\e_{\nu_p}.
\end{equation}
A vector subspace of the algebra spanned by the generators of unit grade is
identified with $\R^D$. Indeed, this is a subspace of $\cl{D-1}$. An element
of this subspace, written
\begin{equation}
\label{xpand1}
x=x^0e_0+x^1e_1+\cdots+x^{D-1}e_{D-1},
\end{equation}
is identified with the point $x=(x^0,x^1,\cdots,x^{D-1})$ of $\R^D$.
This is referred  to as a vector. We use the same notation for vectors
of $\cl{D-1}$ and $\R^D$.
The Clifford algebra $\cl{D}$ is bestowed with a quadratic form that coincides with the Euclidean norm of 
a vector. 

A Clifford algebra possesses a unique canonical automorphism 
$\varrho: \cl{D}\rt\cl{D}$, defined in
terms of the basis elements as
\begin{equation}
\begin{gathered}
\varrho(\e_{\nu})= -\e_{\nu},\\
\varrho(\e_{\nu_1}\e_{\nu_2}\cdots\e_{\nu_p})=
(-1)^p\e_{\nu_1}\e_{\nu_2}\cdots\e_{\nu_p}.
\end{gathered}
\end{equation}
It also possesses a unique canonical anti-automorphism 
$\tau: \cl{D}\rt\cl{D}$, defined by
\begin{equation}
\begin{gathered}
\tau(\e_{\nu})= \e_{\nu},\\
\tau(\e_{\nu_1}\e_{\nu_2}\cdots\e_{\nu_p})=
\e_{\nu_p}\e_{\nu_{p-1}}\cdots\e_{\nu_1}.
\end{gathered}
\end{equation}
These two are commuting involutions
used to define a third involution, called conjugation, in the algebra,
\begin{equation}
\label{bar}
\X\rt\bar{\X}= \tau(\varrho(\X)),
\end{equation}
for every $\X\in\cl{D}$. In particular, 
\begin{equation}
\label{ebar}
\bar{\e}_{\nu}=-\e_{\nu}.
\end{equation}
The coefficient of $e_0$ in $\X$ written in the form \eq{xpand} is called its trace,
denoted $\tr\X$. 
The trace is linear and invariant under cyclic permutations, similarities and 
conjugations \cite{shiro11}. 
For a vector $x$, the trace is 
\begin{equation}
\tr x = \frac{1}{2}(x+\bar{x}).
\end{equation}
The norm-squared of an element of $\cl{D}$ is defined as
\begin{equation}
    \label{sqnorm}
    |\X|^2=\frac{1}{2}(\X\bar{\X} +\bar{\X}\X).
\end{equation}
For a vector $x\in\cl{D-1}$, this matches the Euclidean norm of $x\in\R^D$.
If the norm of a vector $x$  is not zero, then it is called invertible. 
The inverse of a vector $x$ in the Clifford algebra is  
\begin{equation}
\label{qin}
x^{-1} = \frac{\bar{x}}{|x|^2}.
\end{equation}
Invertible vectors form
the multiplicative subgroup $\cls{D}$ of $\cl{D}$.
The subgroup
\begin{equation}
\Ga_D=\{q\in\cls{D-1}|\ \varrho(q)xq^{-1}\in\R^D| \forall x\in\R^D\}
\end{equation}
of $\cl{D-1}$ is called the Clifford group.
Elements of the Clifford group are those obtained by multiplying non-null 
vectors. A matrix $g$ 
\begin{equation}
\label{cliffmat}
g=\left\{
\left.\begin{pmatrix}a&b\\c&d\end{pmatrix}\right|\,
\substack{%
a,b,c,d\ \in\ \Ga_D\cup\{0\}\\
a\tau(b),\ c\tau(d) \ \in \ \R^D\\
\quad a\tau(d) -b\tau(c)=1\phantom{abcd}
}%
\right\}
\end{equation}
is called a Vahlen matrix \cite{Ahlf,loun}.
Vahlen matrices form a group called the Vahlen group, denoted $SL(2,\Ga_D)$, reducing to $SL(2,\R)$ 
for $D=1$, $SL(2,\C)$ for $D=2$ and $SL(2,\H)$ for $D=4$. 
The Vahlen group is the double cover of the M\"obius group, since a Vahlen
matrix and its negative correspond to the same M\"obius transformation, 
\begin{equation}
    \label{MT}
    x\rt (ax+b)(cx+d)^{-1},
\end{equation}
for $x\in\R^D$. 
Under this, the difference between two vectors $x_i$ and $x_j$, denoted 
$x_{ij}=x_i-x_j$, transforms as \cite{Ahlf}
\begin{equation}
    \label{qij}
    x_{ij}\rt \big(\tau(cx_j+d)\big)^{-1}x_{ij}(cx_i+d)^{-1}.
\end{equation}
Its norm-squared transforms as
\begin{equation}
    \label{qsq}
    |x_{ij}|^2\rt |cx_i+d|^{-2}|cx_j+d|^{-2}|x_{ij}|^2.
\end{equation}
It follows from \eqref{qij} that the volume element of $\R^D$ 
transforms as \cite{Ahlf}
\begin{equation}
    \label{vol}
    d^Dx\rt |cx+d|^{-2D}d^Dx.
\end{equation}
For later use, let us introduce the quantities
\begin{equation}
    \label{chi}
    \chi_{ijkl}=x_{ij}x_{ik}^{-1}x_{kl}x_{jl}^{-1},\; i,j,k,l=1,2,\cdots,N,
\end{equation}
for four distinct vectors transforming under \eq{MT} as
\begin{equation}
    \label{chitrans}
\chi_{ijkl}\rt (\tau(cx_j+d))^{-1}\chi_{ijkl}\
\tau(ax_j+b),
\end{equation}
so that the norm-squared
\begin{equation}
    \label{chinorm}
    |\chi_{ijkl}|^2=\chi_{ijkl}\bar{\chi}_{ijkl}=\frac{|x_{ij}|^2|x_{kl}|^2}
{|x_{ik}|^2|x_{jl}|^2}
\end{equation}
is  invariant under the M\"obius transformation \eq{MT}. 
Expanding $\chi$  as an element of $\cl{D-1}$ as \eq{xpand}, 
we derive its trace as
\begin{equation}
    \label{chi0}
    \tr\chi_{ijkl}=\frac{1}{2}(1+|\chi_{ijkl}|^2-|\chi_{lijk}|^2).
\end{equation}
Clearly, this is also invariant under the M\"obius transformation. The derivation of this relation is sketched in appendix \ref{trchi}.

%
\section{The conformal integral}
\label{repmob}
Let us consider $N$ points as the vectors $x_i$ in $\R^D$, $i=1,2,\cdots,N$, 
collectively written as $\bx=(x_1,x_2,\cdots,x_N)$,
in the one-point extension of the Euclidean space,
$\M=\R^D\cup\{\infty\}$. Let us consider the integral
\begin{equation}
    \label{inmu}
    \inmu{\m}{N}{\bx}=\int\frac{d^Dx}{|x-x_1|^{2\mu_1}|x-x_2|^{2\mu_2}\cdots 
|x-x_N|^{2\mu_N}},
\end{equation}
where $\m=(\mu_1,\mu_2,\cdots,\mu_N)$ is an $N$-tuple of real parameters 
called weights, with the total weight, $|\m|=\mu_1+\mu_2+\cdots+\mu_N$. If
$|\m|=D$, then, under simultaneous transformations of the 
the $x_i$'s given by \eq{MT}, $x_i\rt x'_i=(ax_i+b)(cx_i+d)^{-1}$,
the integral $\inmu{\m}{N}{\bx}$ transforms as
\begin{equation}
\inmu{\m}{N}{\bx'}= |cx_1+d|^{2\mu_1} |cx_2+d|^{2\mu_2}
\cdots|cx_N+d|^{2\mu_N}\inmu{\m}{N}{\bx}
\end{equation}
by \eq{qsq} and \eq{vol}. 
The condition $|\m|=D$ is required to ensure that the 
the variable of integration
$x$ remains intact, due to \eq{vol} and \eq{MT}.
The integral $\inmu{\m}{N}{\bx}$ is called a conformal integral. It is 
a representation of the conformal group.
In a specific conformal field theory, the weights are fixed by the conformal
weights of  external points and exchange operators. 
Following the two-dimensional case \cite{looi}, 
we interpret the conformal integral \eq{inmu} as a function on 
the configuration space of $N$ points $\{x_i\}$ on $\M$.  
The space of $N$ non-coincident points in $\M$ is
\begin{equation}
    \label{conf}
\conf_N(\M)=\M^N\setminus\Delta_N
\end{equation}
where 
\begin{equation}
\Delta_N= \{(x_1,x_2,\cdots,x_N)\in\M^N|x_i=x_j \ 
\text{for some}\ i\neq j\}
\end{equation}
is the fat diagonals of $\M^N$. 
As the model of configuration space, we adopt the Fulton-Macpherson completion 
of $\conf_N(\M)$, given by the embedding \cite{sinha,KMV}
\begin{equation}
\label{sinha}
    \begin{gathered}
    \conf_N(\M)\xhookrightarrow{\phantom{xxxxx}}\M^N\times\left(S^{D-1}\right)^{\binom{N}{2}}\times[0,\infty]^{\binom{N}{3}},\\
    (x_1,x_2,\cdots,x_N)\longmapsto (x_1,x_2,\cdots,x_N,v_{12},\cdots,v_{(N-1)N},a_{123},\cdots,a_{(N-2)(N-1)N}),
    \end{gathered}
\end{equation}
where $v_{ij}=\frac{x_{ij}}{|x_{ij}|}$
describes a $(D-1)$-dimensional sphere, $S^{D-1}$, and 
   $a_{ijk}=\frac{|x_{ij}|}{|x_{ik}|}$
are positive real numbers, possibly infinite, for each $i,j,k$.
The group of permutations of the $x_i$'s, \emph{viz.}
$S_N$, acts on the configuration space.  
The quotient of the space $\conf_N(\M)$ by 
$S_N$ is called the configuration space of unordered
points.  Indeed, the conformal 
integral \eq{inmu} is invariant under the permutations, 
provided the weights are permuted accordingly. Oftentimes, 
in different applications, as in the conformal bootstrap programme, 
for example, we need to consider the configuration space with
only a partially ordered set of points, obtained by quotienting $\conf_N(\M)$
with a subgroup of $S_N$.

The configuration spaces $\conf_N(\M)$ and $\conf_{N-1}(\M)$ are related by 
taking the $N$-th point to infinity \cite{sinha}. The conformal integrals
for $N$ and $(N-1)$ points share a similar relation. 
The point $x_N$ in \eq{inmu}
can be taken far by setting $\mu_N=0$, still satisfying $|\m|=D$. We use
this as a check on consistency of expressions later on.

Regular functions on the configuration space
can be expressed in terms of $x_i$, $v_{ij}$ and $a_{ijk}$. 
The  rotation and translation
invariance of the conformal integral imply 
that it can not be a function of $v_{ij}$ or lone $x_i$'s. 
Looked upon as a function on $\conf_N(\M)$, 
germs of $\inmu{\m}{N}{\bx}$ can be expressed in terms of 
\begin{equation}
\label{xa}
|x_{ij}|, a_{ijk}
\end{equation}
only. In view of this, we choose to express the conformal integral 
\eq{inmu} in the form
\cite{PR1,PR2}
\begin{equation}
    \label{i0}
    \inmu{\m}{N}{\bx}=\prod_{%
\substack{i,j\\i<j}}^N|x_{ij}|^{2\beta_{ij}}I_0(\bxi),
\end{equation}
where $\beta_{ij}$'s are  chosen such that
\begin{equation}
\label{betaconstr}
\beta_{ji}=\beta_{ij},\quad\beta_{ii}=0,\quad 
\sum_{j=1}^{N}\beta_{ij}= -\mu_i,
\end{equation}
for each $i,j=1,2,\cdots,N$, 
and $I_0$ is a function of $N_0=N(N-3)/2$ cross ratios,
$\bxi=\{\xi^A| A=1,2,\cdots,N_0\}$. The cross ratios, written as 
\begin{equation}
    \label{inv}
    \xi^A=\prod_{\substack{i,j\\i<j}}^{N}|x_{ij}|^{2\ell^A_{ij}},
\end{equation}
with $\ell$'s taken to be symmetric in the two indices, 
$\ell^A_{ji}=\ell^A_{ij}$, for each $A$, 
$\ell^A_{ii}=0$ for each $A$ and $i$ and are invariant 
under the M\"obius transformation, provided
\begin{equation}
\label{alpha}
\sum_{j=1}^{N}\ell^A_{ij}=0
\end{equation}
for each $A$ and $i=1,2,\cdots,N$.
Let us point out that the cross ratios thus defined can be expressed in terms of the
real numbers $a_{ijk}$.
Differentiating \eq{inmu} with respect to the $x_i$ under the integral sign 
and using the identity
\begin{equation}
    \label{qid}
    (x-x_i)^{-1}x_{ij}(x-x_j)^{-1}=(x-x_i)^{-1}-(x-x_j)^{-1},
\end{equation}
we derive a system of differential equations for the conformal 
integral
\begin{equation}
    \label{lau}
    \sum_{\nu,\nu'=0}^{D-1}\bar{e}_{\nu}x_{ij}\bar{e}_{\nu'}
\frac{\pa^2\inmu{\m}{N}{\bx}}{\pa x_{i}^{\nu}\pa x_{j}^{\nu'}}
=2\sum_{\nu=0}^{D-1}\left(\mu_j\bar{e}_{\nu}\frac{\pa\inmu{\m}{N}{\bx}}{
\pa x_{i}^{\nu}} -\mu_i\bar{e}_{\nu}\frac{\pa\inmu{\m}{N}{\bx}}{\pa x_{j}^{\nu}}
\right),
\end{equation}
generalizing the Lauricella system of the two-dimensional case \cite{looi}. 
Here $x_{i}^{\nu}$ denotes the coefficient of $e_{\nu}$ in $x_i$, as in 
\eq{xpand1}.
Inserting \eq{i0} as an ansatz in it we obtain a system of
differential equations for  
$I_0(\bxi)$. However, the equations contain Clifford numbers $\chi_{ijkl}$.
The invariant equations for the invariant $I_0(\bxi)$ are obtained by 
taking the trace of the equations using \eq{chi0}. The system 
assumes a form similar to that for the quaternionic case treated earlier 
\cite{PR2}, 
\begin{equation}
    \label{Lij}
    \L_{ij}I_0(\bxi)=0,
\end{equation}
with
\begin{equation}
    \label{diffreal}
    \L_{ij}=\sum_{\substack{k,l\\
    1\leq k,l\leq N\\
    k\neq i, l\neq j}}(|\chi_{ijkl}|^2-|\chi_{lijk}|^2)\vartheta_{ik}\vartheta_{jl}+D\vartheta_{ij}+\mu_i\mu_j,
\end{equation}
where we defined 
$\vartheta_{ij}=\sum_{A=1}^{N_0}\ell^A_{ij}\frac{\pa}{\pa\xi^A}
+ \beta_{ij}$.
Using the symmetries of $\L_{ij}$ arising from the identities 
\begin{equation}
    \label{chirel}
    |\chi_{jilk}|^2=|\chi_{ijkl}|^2,\qquad|\chi_{kjil}|^2=|\chi_{lijk}|^2,
\end{equation}
it can be checked that the number of independent differential equations
is  $N_0=N(N-3)/2$, as required for $N_0$ cross ratios. 

The system of equations is amenable to a toric description \cite{PR2}. 
The indices of $x_i$'s as they appear in the 
M\"obius transformation of $|x_{ij}|$ in \eq{qij} are collected in an
$N\times \binom{N}{2}$ matrix $\mA$, which takes the form
\begin{equation}
\label{Adef}
\mA_{i,jk}=\delta_{ij}+\delta_{ik}.
\end{equation}
Let us point out that the formulation of the conformal group in terms of the
Vahlen group is cricial in writing the conformal transformations as
M\"obius transformations. It makes the toric matrix conspicuous, which 
is can not be dones using Ecuclidean vectors. 
Let $\ell^A_{ij}$, labelled by $A$, 
be $N_0$ integer vectors annihilated by $\mA$,
\begin{equation}
\label{AL0}
\sum_{\left\{\substack{(jk)|j<k\\j,k=1,2,\cdots,{N}}\right\}}
\mA_{i,jk}\ell^A_{jk}=0.
\end{equation}
Arranging these vectors along rows yields an $N_0\times\binom{N}{2}$ matrix
$L$, which we refer to as the Gale matrix. The GKZ toric ideal is then 
generated by 
\begin{gather}
    \label{GKZ1}
    \sum_{\substack{j,k=1\\j<k}}^N \mA_{i,jk} 
|x_{jk}|^2\partial_{jk} +\mu_i,\quad \forall i,\\
\label{GKZ2}
\prod_{\ell^A_{ij}>0} \partial_{ij}^{\ell^A_{ij}} 
-\prod_{\ell^A_{ij}<0} \partial_{ij}^{-\ell^A_{ij}},
\ A=1,2,\cdots,N_0,
\end{gather}
where $\partial_{ij}=\frac{\partial}{\partial |x_{ij}|^2}$. 
The rows of the Gale matrix thus define invariants of the M\"obius
transformation, that is, the cross ratios. 
Thus, the $\ell^A_{ij}$ are integers, and as is conspicuous from the 
notation, are identified with the exponents in \eq{inv}, satisfying 
\eq{alpha}. 

The set of differential equations ensuing from the GKZ ideal 
are solved by the simultaneous solution of the equations
\begin{equation}
    \label{LI0}
    \L_{ijkl}I_0(\bxi)=0,
\end{equation}
where
\begin{equation}
    \label{Lijkl}
    \L_{ijkl}=\vartheta_{ij}\vartheta_{kl}-|\chi_{ijkl}|^2\vartheta_{ik}\vartheta_{jl}.
\end{equation}
We refer to these as the GKZ system.
Observing the relation 
\begin{equation}
    \label{LL}
    \L_{ij}=\sum_{\substack{k=1\\ k\neq i,j}}^N\sum_{\substack{l=1\\ l\neq i,j,k}}^N\left(\L_{lijk}-\L_{ijkl}\right),
\end{equation}
we deduce that the same solution solves \eq{Lij}, thereby giving the invariant
function $I_0(\bxi)$. 
These observations allow writing the differential equations for
the conformal integral \eq{i0} directly as
\begin{equation}
\hat{\L}_{ijkl} \inmu{\m}{N}{\bx} =0,
\end{equation}
with $\hat{\L}_{ijkl}=\pa_{ij}\pa_{kl}-\pa_{ik}\pa_{jl}$. 
Hence, the conformal 
integral $\inmu{\m}{N}{\bx}$ is obtained directly as a solution to the GKZ
system. Let us point out that in $D=2$ there exists another way to 
associate a GKZ system to a conformal integral \cite{stien}, but it 
fixes one of the points, which we avoid. 

We now set out to derive an expression for the conformal integral 
by solving the GKZ system \cite{mheo}. Let 
$\Sigma=\{1,2,\cdots,\binom{N}{2}\}$.
Let the columns of the toric matrix $\mA$ defined in
\eq{Adef} be denoted by $\ba_I$, $I\in\Sigma$.
Let us point out that a column of $\mA$, in the notation of \eq{Adef} is 
labelled by a bi-index, $jk$. Thus, $I=jk,\ j<k$ for some $j$ and $k$. 
However, for the rest of this section, in order to not clutter the notation
too much,  we use a single index, a capital Roman 
letter, reverting to the bi-indices from the next section. 
Each of the columns $\ba_I$ is an $N$-dimensional real vector.
Let $\sigma$ be a subset of $\Sigma$. The positive span of the column vectors,
indexed by $\sigma$ is called a cone, namely,
\begin{equation}
\cone{\sigma} = \sum_{I\in\sigma} \R_{\geqslant 0}\ba_I.
\end{equation}
The subset $\sigma$, the subset of vectors $\{\ba_I|\ I\in\sigma\}$ and
$\cone{\sigma}$ are used to refer to each other interchangeably. 
The complement of $\sigma$ in $\Sigma$ is denoted $\bsigma$.
A subset $\mathcal{T}$ of the power set of $\Sigma$ is called a 
triangulation if 
$\{\cone{\sigma}|\sigma\in \mathcal{T}\}$ is the set of cones in 
a simplicial fan with 
support equal to $\cone{\mA}$. For an $N$-dimensional real vector $w$, 
a triangulation $\mathcal{T}(w)$ is defined if there exists an 
$N$-dimensional real vector $\mathbf{m}$ such that  
\begin{equation}
\mathbf{m}\cdot\ba_I  \begin{cases}
=w_I,\quad \text{if}\ I\in\sigma,\\
< w_I, \quad\text{if}\ I\in\bsigma.
\end{cases}
\end{equation}
A triangulation is regular if $\mathcal{T}=\mathcal{T}(w)$ for some $w$.

Given a regular triangulation $\mathcal{T}$ 
corresponding to the columns of $\mA$, let 
$\sA$ denote the $N\times N$ matrix made of $\{\ba_I\}$, $I\in\sigma$ and 
$\bA$ denote the $N\times N_0$ matrix made of the vectors $\{\ba_I\}$, 
$I\in\bsigma$. By rearranging the columns, as required, the toric  matrix
$\mA$ can be expressed as $\mA=(\sA|\bA)$ with the columns reshuffled.
A formal solution of the GKZ system \eq{GKZ1} and $\eq{GKZ2}$ is then 
given by the  $\Gamma$-series, 
\begin{equation}
\label{G-ser}
\Psi_{\sigma} = 
\prod_{I\in\sigma}
|x_{I}|^{-\left(\sA^{-1}\m\right)_I}
\sum_{\mathbf{n}\in\Z_{\geq 0}^{N_0}} \frac{%
\prod_{I\in\sigma}
|x_I|^{-\left(\sA^{-1}\bA\mathbf{n}\right)_I}
\prod_{J\in\bsigma}
|x_J|^{{n}_J}}%
{%
\prod_{I\in\sigma}
\G{1-\left(\sA^{-1}\m\right)_I-\left(\sA^{-1}\bA\mathbf{n}\right)_I}
\prod_{J=1}^{N_0}\G{1+n_J}
},
\end{equation}
where $(\mathbf{v})_I={v}_I$ is taken to denote the $I$-th components 
of a vector $\mathbf{v}$ and 
$x_I$ denotes the appropriately identified variable $x_{ij}$ introduced 
above.

Given a regular triangulation $\mathcal{T}=\{\sigma_i\}$, the collection of 
$\Gamma$-series corresponding to
its cones, $\{\Psi_i=\Psi_{\sigma_i}\}$,
provides a basis of solutions of the GKZ system off the
discriminat locus. For $N>3$, these can be expressed in terms 
of the cross-ratios. The conformal integral \eq{inmu} is a linear combination
of these $\Gamma$-series
\begin{equation}
\inmu{\m}{N}{\bx} = \sum_{\stackrel{i}{\sigma_i\in\mathcal{T}}} C_i\Psi_i.
\end{equation}
The coefficients 
may be chosen so as to make the solution invariant under the permutation
symmetry of the conformal integral. 
From \eq{Adef} we note that a permutation of the points $x_i$ corresponds to 
shuffling the columns of the toric matrix $\mathcal{A}$. This, in turn, alters
the definition of cross ratios \eq{inv} and corresponds to analytic
continuation of the $\Gamma$-series solutions $\Psi_{\sigma}$. 
\comment{
The new set of solutions is related to the set of old ones by monodromy. 
From the examples it turns out that the
fully symmetric solution of \eq{inmu} on the unordered configuration space 
is given by the constants
\begin{equation}
\label{cigma}
C_i=\frac{1}{\prod_{I\in\sigma_i}\sin\pi\!\left({A^{-1}_{\sigma_i}}
\m\right)_I}. 
\end{equation}
corresponding to the cone $\sigma_i$.
These are the same as those appearing in the expression for the unique 
monodromy-invariant Hermitian form \cite{beu,mheo2,ver}.
For generic values of the weights $\m$ the GKZ system being irreducible 
\cite{SW} with 
holonomic rank greater than unity
appears to contradict the fact that there is no monodromy-invariant solution
\footnote{We thank the anonymous referee for pointing this out to us, 
along with ref~\cite{SW}, which
lead to the clarification of the results}.
The resolution to this appararent paradox lies in the fact that the permutation
symmetry of \eq{inmu} permutes the $x_i$ and $\mu_i$ together. 
Strictly speaking, one is considering  different GKZ systems related by the 
permutations. The solutions are to be compared after analytic continuation 
as demonstrated in \cite{PR2} for $N=4$.
}
\section{Examples}
\label{ex456}
Let us present the expressions for conformal integrals as 
solutions of the GKZ system in some examples.
\subsection{$N=2$ \& $N=3$ points}
Let us start with the well-known cases of $N=2$ and $N=3$ points for
completeness. Since the number of cross ratios is $N_0=N(N-3)/2$,
defining a non-trivial cross ratio requires at least four points. The 
conformal integral for $N=2$ and $N=3$ points 
are completely fixed by the conformal 
symmetry. This is reflected in the GKZ system by the absence of \eq{GKZ2} in 
these cases. The integrals are determined by \eq{GKZ1}.

For $N=2$, $\mA=\left(\begin{smallmatrix}1\\1\end{smallmatrix}\right)$ with
a trivial kernel.
Equation \eq{GKZ1} can be solved only for $\mu_1=\mu_2$, which, due to the
constraint $\mu_1+\mu_2=D$ means $\mu_1=\mu_2=D/2$. The equation takes the form
\begin{equation}
\left(|x_{12}|^2\pa_{12} + D/2\right)\inmu{\m}{2}{\bx} =0,
\end{equation}
which is solved as 
\begin{equation}
\inmu{\m}{2}{\bx} = \frac{1}{|x_{12}|^{D}}
\end{equation}
In the case of $N=3$, the toric matrix is
\begin{equation}
        \label{amat3}
\mA=\bordermatrix{%
 & \scriptstyle (12) &\scriptstyle  (13) &\scriptstyle  (23) \cr
 & 1&1&0\cr
 & 1&0&1\cr
 & 0&1&1\cr
}.%
\end{equation}
Again, the kernel is trivial, equation \eq{GKZ2} is absent. The equations
\eq{GKZ1} take the form
\begin{equation}
\label{gkz3}
\begin{split}
\left(|x_{12}|^2\pa_{12} +|x_{13}|^2\pa_{13}
+\mu_1\right)\inmu{\m}{3}{\bx} &=0\\
\left(|x_{12}|^2\pa_{12} +|x_{23}|^2\pa_{23}
+\mu_2\right)\inmu{\m}{3}{\bx} &=0\\
\left(|x_{13}|^2\pa_{13} +|x_{23}|^2\pa_{23}
+\mu_3\right)\inmu{\m}{3}{\bx} &=0.
\end{split}
\end{equation}
The unique solution to these equations is 
\begin{equation}
\label{psi3}
\inmu{\m}{3}{\bx} = 
|x_{12}|^{\mu_3-\mu_1-\mu_2}
|x_{13}|^{\mu_2-\mu_1-\mu_3}
|x_{23}|^{\mu_1-\mu_2-\mu_3},
\end{equation}
as expected. Let us stress that, while the three equations \eq{gkz3}
are solved without 
any restriction on the $\mu$'s, the very derivation of the equations was 
performed with the assumtion $\mu_1+\mu_2+\mu_3=D$.
No invariant cross ratios appear in these two cases.
\subsection{$N=4$ points}
In this and the following examples the toric matrices have non-trivial kernels,
leading to non-trivial cross ratios. Hence both \eq{GKZ1} and \eq{GKZ2}
are to be solved. For these cases \eq{GKZ1} is taken into account through
the restriction \eq{betaconstr} on the $\beta$'s appearing in the 
ansatz \eq{i0}.

The toric matrix in the case of $N=4$ is given by \cite{PR2}
\begin{equation}
        \label{amat4}
\mA=\bordermatrix{%
\scriptstyle\stackrel{I}{(ij)}&\scriptstyle \stackrel{1}{(12)} 
&\scriptstyle\stackrel{2}{(13)}
&\scriptstyle\stackrel{3}{(14)} 
&\scriptstyle\stackrel{4}{(23)}
&\scriptstyle\stackrel{5}{(24)} 
&\scriptstyle \stackrel{6}{(34)}\cr
 & 1&1&1&0&0&0\cr
 & 1&0&0&1&1&0\cr
 & 0&1&0&1&0&1\cr
 & 0&0&1&0&1&1\cr
}.%
\end{equation}
We have indicated the two ways of labelling the columns that has been used. 
The Gale matrix, whose elements are $\ell_{ij}^A$ defined in \eq{inv},
is chosen as the basis of generators of the Gr\"obner basis of the
GKZ ideal \footnote{The choice of basis of $L$ 
is different from that in \cite{PR2}. Hence the choice of cross ratios are
also different.}
\begin{equation}
        \label{v4}
L= \bordermatrix{%
&&&&&\cr
\bell^1 &0&1&-1&-1&1&0\cr
\bell^2 & 1&0&-1&-1&0&1
}\ %
\begin{matrix}
\xi^1\\\xi^2
\end{matrix}
\end{equation}
The rows of the Gale matrix \eq{v4} give rise to the
two cross ratios,
\begin{equation}
\label{xrat4}
\xi^1=\frac{|x_{13}|^2|x_{24}|^2}{|x_{14}|^2|x_{23}|^2},\quad
\xi^2=\frac{|x_{12}|^2|x_{34}|^2}{|x_{14}|^2|x_{23}|^2}.
\end{equation}  
The connected component of the 
discriminant locus, derived from the toric data \eq{amat4} is 
\begin{equation}
\label{disc4}
1-2(\xi^1+\xi^2)+(\xi^1-\xi^2)^2=0,
\end{equation}
or, equivalently,
\begin{equation}
\sqrt{\xi^1} +\sqrt{\xi^2}=1.
\end{equation}
The GKZ system \eq{GKZ1} and \eq{GKZ2} has holonomic rank $4$, which is the
rank of the corresponding $\mathcal{D}$-module. 
Accordingly, a regular triangulation is given by 
\begin{equation}
\label{T4}
\mathcal{T} = \left\{ \{1,2,3,4\},\{1,3,4,5\},\{2,3,4,6\},\{3,4,5,6\}\right\},
\end{equation}
or, equaivalently, by 
\begin{equation}
\mathcal{T} = \scriptstyle
\left\{ \{(12),(13),(14),(23)\},\{(12),(14),(23),(24)\},
\{(13),(14),(23),(34)\},\{(14),(23),(24),(34)\}\right\},
\end{equation}
in the two labelling schemes. It contains four cones
with height vector 
$w=(\tfrac{1}{2},\tfrac{1}{2},0,0,0,0)^T$.
There are, therefore, four linearly independent solutions
to the GKZ system, obtained from \eq{G-ser}.
\begin{multline}
\label{psi41}
\{1,2,3,4\}:\quad 
\Psi^{(4)}_1=
|x_{12}|^{D-2(\mu_1+\mu_2)}
|x_{13}|^{D-2(\mu_1+\mu_3)}
|x_{14}|^{-2\mu_4}
|x_{23}|^{2\mu_1-D} \\
\sum_{n_1,n_2=0}^{\infty}
\frac{(\xi^1)^{n_1}(\xi^2)^{n_2}}{\scriptstyle%
\G{1+n_1}
\G{1+n_2}
\G{1+n_1-\mu_1-\mu_3+D/2}
\G{1+n_2-\mu_1-\mu_2+D/2}
\G{1-n_1-n_2-\mu_4}
\G{1-n_1-n_2+\mu_1-D/2)}
},
\end{multline}
\begin{multline}
\label{psi42}
\{ 1,3,4,5\}:\quad 
\Psi^{(4)}_2=
|x_{12}|^{D-2(\mu_1+\mu_2)}
|x_{14}|^{2\mu_2-D}
|x_{23}|^{-2\mu_3}
|x_{24}|^{D-2(\mu_2+\mu_4)} \\
\sum_{n_1,n_2=0}^{\infty}
\frac{(\xi^1)^{n_1}(\xi^2)^{n_2}}{\scriptstyle%
\G{1+n_1}
\G{1+n_2}
\G{1+n_1-\mu_2-\mu_4+D/2}
\G{1+n_2-\mu_1-\mu_2+D/2}
\G{1-n_1-n_2-\mu_3}
\G{1-n_1-n_2+\mu_2-D/2}
},
\end{multline}
\begin{multline}
\{2,3,4,6\}:\quad 
\label{psi43}
\Psi^{(4)}_3=
|x_{13}|^{D-2(\mu_1+\mu_3)}
|x_{14}|^{2\mu_3-D}
|x_{23}|^{-2\mu_2}
|x_{34}|^{D-2(\mu_3+\mu_4)}\times \\
\sum_{n_1,n_2=0}^{\infty}
\frac{(\xi^1)^{n_1}(\xi^2)^{n_2}}{\scriptstyle%
\G{1+n_1}
\G{1+n_2}
\G{1+n_1-\mu_1-\mu_3+D/2}
\G{1+n_2-\mu_3-\mu_4+D/2}
\G{1-n_1-n_2-\mu_2}
\G{1-n_1-n_2+\mu_3-D/2}
},
\end{multline}
\begin{multline}
\label{psi44}
\{3,4,5,6\}:\quad 
\Psi^{(4)}_4=
|x_{14}|^{-2\mu_1}
|x_{23}|^{2\mu_4-D}
|x_{24}|^{D-2(\mu_2+\mu_4)}
|x_{34}|^{D-2(\mu_3+\mu_4)} \times \\
\sum_{n_1,n_2=0}^{\infty}
\frac{(\xi^1)^{n_1}(\xi^2)^{n_2}}{\scriptstyle%
\G{1+n_1}\G{n_2+1}
\G{1+n_1-\mu_2-\mu_4+D/2}
\G{1+n_2-\mu_3-\mu_4+D/2}
\G{1-n_1-n_2-\mu_1}
\G{1-n_1-n_2+\mu_4-D/2}
},
\end{multline}
where the constraint $\mu_1+\mu_2+\mu_3+\mu_4=D$ has been used. 
We have indicated the cones to which the solutions correspond.
These are the same as in the four dimensional case, with 
the dimension changed to $D$ \cite{PR1,PR2}.
The conformal integral $\inmu{\m}{4}{\bx}$ is 
a linear combination of these with constant coefficients,
\begin{equation}
    I_4^{\m} =C_1(\m)\Psi^{(4)}_1+C_2(\m)\Psi^{(4)}_2
+C_3(\m)\Psi^{(4)}_3+C_4(\m)\Psi^{(4)}_4.
\end{equation}
The constants can be fixed from asymptotic or boundary conditions, or by
imposing permutation symmetry $S_4$ of the four points.
The four solutions can be expressed in terms of the Appell function $F_4$. 
Using various formul{\ae} for change of
variables of the Appell functions \cite{DO3,PR2} corresponding
to analytic continuation of the series, the constants are fixed as
\begin{equation}
\begin{split}
C_1(\m)^{-1} &= \sin\pi{\mu_4} \sin\pi\!\left({\frac{D}{2}-\mu_1}\right) 
\sin\pi\!\left({\frac{D}{2}-\mu_2-\mu_4}\right)
\sin\pi\!\left({\frac{D}{2}-\mu_3-\mu_4}\right),\\
C_2(\m)^{-1} &= \sin\pi{\mu_3} \sin\pi\!\left({\frac{D}{2}-\mu_2}\right)
\sin\pi\!\left({\frac{D}{2}-\mu_1-\mu_3}\right)
\sin\pi\!\left({\frac{D}{2}-\mu_3-\mu_4}\right),\\
C_3(\m)^{-1} &= \sin\pi{\mu_2} \sin\pi\!\left({\frac{D}{2}-\mu_3}\right)
\sin\pi\!\left({\frac{D}{2}-\mu_1-\mu_2}\right)
\sin\pi\!\left({\frac{D}{2}-\mu_2-\mu_4}\right),\\
C_4(\m)^{-1} &= \sin\pi{\mu_1}\sin\pi\!\left({\frac{D}{2}-\mu_4}\right)
\sin\pi\!\left({\frac{D}{2}-\mu_1-\mu_2}\right)
\sin\pi\!\left({\frac{D}{2}-\mu_1-\mu_3}\right),
\end{split}
\end{equation} 
as in \eq{cigma}. 

Let us now discuss the reduction of points. As discussed in 
section~\ref{repmob}, taking $x_4$ to infinity and setting $\mu_4=0$, the 
four-point conformal integral reduces to the three-point integral. Putting
$\mu_4=0$ in \eq{psi41} gets rid of $x_{14}$, leaving only three points. Then,
the factor $\G{1-n_1-n_2-\mu_4}$ in the denominator forces $n_1=n_2=0$, since
for other values of the integers the series vanishes due to
the singularity of the Gamma function. We obtain \eq{psi3}.
However, in this limit the existence of the other three solutions  is not
guaranteed. In order for those to exist, we need to fix the indices in such
a way that the index of $|x_{i4}|$ is zero for each $i=1,2,3$. 
For $\Psi^{(4)}_2$, for example, this means, we have to take 
$2\mu_2-D=0$. Then, the last factor in the denominator restricts 
the integers to $n_1=n_2=0$. The solution reduces to \eq{psi3} with 
$\mu_1+\mu_3=\mu_2$, equivalent to choosing $2\mu_2=D=\mu_1+\mu_2+\mu_3$. 
Other solutions reduce similarly to the three-point  integral, but for special 
values of the weights.
\subsection{$N=5$ points}
For five points, $N_0=5$ and $\binom{N}{2}=10$. The toric matrix is 
\begin{equation}
\label{amat5}
\mA=\bordermatrix{%
\scriptstyle\stackrel{I}{(ij)}
&\scriptstyle\stackrel{1}{(12)} &\scriptstyle\stackrel{2}{(13)}
&\scriptstyle\stackrel{3}{(14)} &\scriptstyle\stackrel{4}{(15)}
&\scriptstyle\stackrel{5}{(23)} &\scriptstyle\stackrel{6}{(24)}
&\scriptstyle\stackrel{7}{(25)} &\scriptstyle\stackrel{8}{(34)} 
&\scriptstyle\stackrel{9}{(35)} &\scriptstyle\stackrel{10}{(45)} \cr
& 1&1&1&1&0&0&0&0&0&0\cr
& 1&0&0&0&1&1&1&0&0&0\cr
& 0&1&0&0&1&0&0&1&1&0\cr
& 0&0&1&0&0&1&0&1&0&1\cr
& 0&0&0&1&0&0&1&0&1&1\cr
}.%
\end{equation} 
The Gale matrix is chosen as the basis of generators of the Gr\"obner basis 
of the GKZ ideal as
\footnote{This choice is also different from that in \cite{PR2}}
\begin{equation}
\label{v5}
L=\bordermatrix{%
&&&&&&&&&&\cr
\bell^1& 0&1&-1&0&-1&1&0&0&0&0\cr
\bell^2& 0&1&0&-1&-1&0&1&0&0&0\cr
\bell^3& 1&0&-1&0&-1&0&0&1&0&0\cr
\bell^4& 1&0&0&-1&-1&0&0&0&1&0\cr
\bell^5& 1&1&-1&-1&-1&0&0&0&0&1\cr
}\ %
\begin{matrix}
\xi^1\\\xi^2\\\xi^3\\\xi^4\\\xi^5
\end{matrix}.
\end{equation} 
The rows of the Gale matrix correspond to the cross ratios
\begin{equation}
\begin{gathered}
\label{xrat5}
\xi^1=\frac{|x_{13}|^2|x_{24}|^2}{|x_{14}|^2|x_{23}|^2},\quad
\xi^2=\frac{|x_{13}|^2|x_{25}|^2}{|x_{15}|^2|x_{23}|^2},\quad
\xi^3=\frac{|x_{12}|^2|x_{34}|^2}{|x_{14}|^2|x_{23}|^2},\\
\xi^4=\frac{|x_{12}|^2|x_{35}|^2}{|x_{15}|^2|x_{23}|^2},\quad
\xi^5=\frac{|x_{12}|^2|x_{13}|^2|x_{45}|^2}{|x_{14}|^2|x_{15}|^2|x_{23}|^2}.
\end{gathered}
\end{equation}
The connected component of the
discriminant locus obtained from the toric data \eq{amat5} is
\begin{multline}
\label{disc5}
\xi^4 (\xi^1)^2 + (\xi^2)^2 \xi^3
-\big((\xi^2+\xi^5+1)\ \xi^4
+\xi^3 (\xi^2+\xi^4-1)-(\xi^2-1)\ \xi^5-(\xi^4)^2\big)\
   \xi^1+\\  
+\xi^5 \big(\xi^3 (\xi^4-1)-\xi^4+\xi^5+1\big)
+\xi^2 \big((\xi^3)^2-(\xi^4+\xi^5+1)
   \xi^3+\xi^4-\xi^5\big)=0.
\end{multline}
The holonomic rank of the GKZ system for $N=5$ is $11$. A regular 
triangulation, out of $102$ possible ones, with $11$ cones is 
\begin{multline}
\label{T5}
\mathcal{T} = \left\{
\{1, 2, 3, 4, 5\},\{1, 3, 4, 5, 6\}, \{1, 4, 5, 6, 7\}, \{2, 3, 4, 5, 8\},
\{3, 4, 5, 6, 8\}, \{4, 5, 6, 7, 8\},\right.
\\ \left.  \{2, 4, 5, 8, 9\}, \{4, 5, 7, 8, 9\},
\{3, 4, 6, 8, 10\},\{4, 6, 7, 8, 10\},\{4, 7, 8, 9, 10\}
\right\},
\end{multline}
with  height vector
 $w=(2,\tfrac{3}{2},1,0,\tfrac{1}{2},\tfrac{1}{2},0,0,0,0)^T$.
Correspondingly, there are $11$ solutions. Again, writing the $\Gamma$-series 
as Mellin-Barnes integrals and using analytic continuation of the 
cross-ratios, we obtain the constants to make the linear combination of these
eleven $\Gamma$-series symmetric under $S_5$. The eleven solutions and 
the corresponding constants, according to
\eq{cigma}, are listed in the appendix \ref{5sol}.

The reduction to the solutions to four points proceed similarly as the
reduction from four to three points, discussed earlier. Upon taking $x_5$ to 
infinity and setting $\mu_5=0$ reduces 
$\Psi^{(5)}_1$, $\Psi^{(5)}_2$, $\Psi^{(5)}_4$ and $\Psi^{(5)}_5$ to the four
solutions \eq{psi41} -- \eq{psi44}, with two cross ratios 
$\xi^1$ and $\xi^3$. That these are the correct ones can be verified by
comparing \eq{xrat4} and \eq{xrat5} --- the other three 
do not yield cross ratios in this limit and do not survive the limit to
contribute to the four-point function.
All the other seven solutions reduce 
to the four-point cases with special values of the weights. 
\section{Discussions and summary}
\label{summ}
An imperfection in the treatment presented here must be mentioned. 
On the face of it, the conformal integrals being expressed 
in terms of the same GKZ hypergeometric functions in any dimension, with only
the dimension changing in the expressions
is surprising. However, let us point out that we have 
considered $N(N-3)/2$ cross ratios in all dimensions from a 
combinatorial estimate. The actual number of independent cross ratios 
is $\operatorname{min}\left(ND-(D+1)(D+2)/2,N(N-3)/2\right)$. Hence 
for sufficiently large $N$ the cross ratios considered here will be 
related. For example, in four dimensions, 
these cross ratios will be related beyond $N=6$. 
The relations, which generalize the Pl\"ucker relations in 
two dimensions, are rather difficult to find in practice.
In view of the 
computational difficulty of the conformal integrals, it seems that imposing 
the relations among the cross ratios second to obtaining the integrals
as GKZ hypergeometric functions will be a satisfactory  strategy. 
We hope the results reported here will  be useful in various 
fields where conformal integrals make an appearance. 

To summarise, we evaluate conformal integrals in any dimension, for any number
of points. The expresions for the cases $N=4,5$ are presentted, but the 
considerations in this article are completely general, subject to the 
limitations described above. We utilise a special formulation \cite{Ahlf,loun} 
to write the M\"obius transformation in $\R^D$ in terms of the
Clifford algebra $Cl_D(\R)$. This formulation is then used to
write the $N$-point conformal integral \eq{inmu} which transforms under the
M\"obius transformation in terms of the elements of $Cl_D(\R)$.  
We interpret the conformal integral
as a representation of the (double cover of the) conformal group, 
written as $SL(2,\Gamma_D)$, on the 
Fulton-Macpherson compactification of the configuration space
of $N$ points on $\R^D$. A set of differential 
equations are then derived by differentiating with respect 
to the positions of the $N$ points. 
The equations are shown to be solved
by GKZ hypergeometric functions corresponding 
to the toric data obtained from the weights of transformation
of the conformal integral under the M\"obius transformation of the 
$N$ points. The formulation in terms of the Clifford algebra is indispensable 
in the sense that it allows writing the
conformal transformations as M\"obius transformations, which is crucial 
in identifying exponents which are collected to write the toric matrix 
\eq{Adef}. The computation of conformal blocks in conformal
field theories, wherein conformal 
integrals appear, utilises Casimir operators of the confomal group. While
we have not proved it in general, it appears that the Casimir operators,
quadratic and higher ones, are within the GKZ $\mathcal{D}$-module and can be expressed
in terms of the differential operators $\hat{\L}_{ijkl}$ considered here. 

We present series solutions for $N=4,5$, called $\Gamma$-series. 
These are the germs of the 
conformal integral. They correspond to regular triangulations of 
the polytope subtended by the columns of the toric matrix \eq{Adef}. 
For $N=4$, the 
solutions can be expressed in terms of the Appell function $F_4$. Generally,
they are Lauricella functions. 
We relate higher point integrals to lower ones by taking limits properly. 
\comment{
Moreover, the conformal integral \eq{inmu} is invariant under permutations
of $x_i$ and $\mu_i$ at once, thereby being defined on the unordered
configuration space. The coefficients appearing in the linear combinations
coincide with the coefficients of the monodromy-invariant Hermitian form, 
based on the examples at hand.
However, it would be more satisfying to
obtain a rigorous derivation of the coefficients. 
}
The choice of the triangulation, used to obtain the 
solutions, breaks this symmetry. Relating the permutations to monodromy
of the $\Gamma$-series solutions we use the unique Hermitian form to write 
down the fully permutation symmetric expression of the conformal integral.
The procedure is general, and can be 
implemented on a computer. 
The triangulations were obtained using \verb|Macaulay2|\footnote{
KR thankfully acknowledges help derived  from the
{\tt Macaulay2} users' group.}, 
while the series solutions associated
to the cones were obtained using \verb|Mathematica|.
The considerations in the present article, restricted to the Euclidean
metric on $\R^D$, generalize to metrics of Lorentzian signature, assuming 
that the choice of variables \eq{xa} is valid.

\begin{appendices}
\setcounter{equation}{0}
\renewcommand{\theequation}{\thesection.\arabic{equation}}
\section{Calculation of trace of $\chi_{ijkl}$}
\label{trchi}
In this appendix, we outline the derivation of \eq{chi0}.
Writing \eq{chi} as
\begin{equation}
\label{psichi}
\chi_{ijkl}=\frac{x_{ij}\bar{x}_{ik}x_{kl}\bar{x}_{jl}%
}{|x_{ik}|^2|x_{jl}|^2},
\end{equation}
and $x_{ik}=x_{ij}+x_{jk}$ and $x_{kl}=x_{jl}-x_{kl}$, the numerator
becomes 
\begin{equation}
\label{psi1}
|x_{ij}|^2x_{kl}\bar{x}_{jl}
+|x_{jl}|^2x_{ij}\bar{x}_{jk}
-|x_{jk}|^2x_{ij}\bar{x}_{jl}.
\end{equation}
For two vectors $x$ and $y$ in $\R^D$,
\begin{equation}
\label{bivec}
x\bar{y}=(x\cdot y)e_0+
\sum_{\nu=1}^{D-1}(x^{\nu}y^0-x^0y^{\nu})e_\nu
-\sum_{\substack{\nu,\nu'=1\\ \nu<\nu'}}^{D-1}(x^\nu y^{\nu'}-x^{\nu'} y^{\nu})e_{\nu}e_{\nu'},
\end{equation}
where $x\cdot y$ denotes the inner product of vectors in the Euclidean
metric of $\R^D$,
namely, $x\cdot y=x^0y^0+x^1y^1+\cdots+x^{D-1}y^{D-1}$. 
Taking trace, that is the coefficient of $e_0$, 
\begin{equation}
\label{trbivec}
\tr{(x\bar{y})}=x\cdot y.
\end{equation}
The trace of the numerator \eq{psi1} becomes
\begin{equation}
\label{trpsi}
|x_{ij}|^2(x_{kl}\cdot {x}_{jl})
+|x_{jl}|^2(x_{ij}\cdot {x}_{jk})
-|x_{jk}|^2(x_{ij}\cdot {x}_{jl}).
\end{equation}
Using the identities
\begin{equation}
\label{pdtiden}
\begin{gathered}
x_{kl}\cdot x_{jl}=\frac{1}{2}(|x_{jl}|^2+|x_{kl}|^2-|x_{jk}|^2),\\
x_{ij}\cdot x_{jk}=\frac{1}{2}(|x_{ik}|^2-|x_{ij}|^2-|x_{jk}|^2),\\
x_{ij}\cdot x_{jl}=\frac{1}{2}(|x_{il}|^2-|x_{ij}|^2-|x_{jl}|^2)
\end{gathered}
\end{equation}
in \eq{trpsi} and simplifying the trace of the numerator 
assumes the form  
\begin{equation}
\label{trpsi1}
\frac{1}{2}(|x_{ik}|^2|x_{jl}|^2+|x_{ij}|^2|x_{kl}|^2-|x_{il}|^2|x_{jk}|^2).
\end{equation}
Plugging this in \eq{psichi} and using \eq{chinorm} we derive \eq{chi0}.
\clearpage
\section{Solutions for $N=5$}
\setcounter{equation}{0}
\label{5sol}
Here we list the $11$ solutions for the case of $N=5$ points. We use the
notation $\mathbf{n}=(n_1,n_2,n_3,n_4,n_5)$, each running over the non-negative 
integers. The cones are indicated for each.
\begin{multline}
\label{psi51}
\{ 1,2,3,4,5\}:\quad 
\Psi_1^{(5)} = 
|x_{12}|^{D-2(\mu_1+\mu_2)} 
|x_{13}|^{D-2(\mu_1+\mu_3)} 
|x_{14}|^{-2 \mu_4} 
|x_{15}|^{-2\mu_5} 
|x_{23}|^{2\mu_1-D}
\\
\sum_{\mathbf{n}\in\Z_{\geqslant 0}^5} 
\frac{(\xi^1)^{n_1} 
(\xi^2)^{n_2} 
(\xi^3)^{n_3} 
(\xi^4)^{n_4} 
(\xi^5)^{n_5}}{\G{1+n_1} \G{1+n_2} \G{1+n_3} \G{1+n_4} \G{1+n_5}}
\\
\frac{1}{\G{1-n_1-n_3-n_5-\mu_4} \G{1-n_2-n_4-n_5-\mu_5}}\\
\frac{1}{\G{1-n_1-n_2-n_3-n_4-n_5+\mu_1-D/2}}
\frac{1}{\G{1+n_1+n_2+n_5-\mu_1-\mu_3+D/2}}\\
\frac{1}{\G{1+n_3+n_4+n_5-\mu_1-\mu_2+D/2}}
\end{multline}
\begin{multline}
\label{psi52}
\{ 1,3,4,5,6\}:\quad
\Psi_2^{(5)}=
|x_{12}|^{D-2(\mu_1+\mu_2)} 
|x_{14}|^{2(\mu_2+\mu_5)-D} 
|x_{15}|^{-2 \mu_5} 
|x_{23}|^{-2 \mu_3} 
|x_{24}|^{2(\mu_1+\mu_3)-D}
\\
\sum_{\mathbf{n}\in\Z_{\geqslant 0}^5} 
\frac{(\xi^1)^{n_1} 
(\xi^2/\xi^1)^{n_2} 
(\xi^3)^{n_3} 
(\xi^4)^{n_4} 
(\xi^5/\xi^1)^{n_5}}{\G{1+n_1} \G{1+n_2} \G{1+n_3} \G{1+n_4} \G{1+n_5}}
\\
\frac{1}{\G{1-n_1-n_3-n_4-\mu_3} \G{1-n_2-n_4-n_5-\mu_5}}\\
\frac{1}{\G{1+n_1-n_2-n_5+\mu_1+\mu_3-D/2}}
\frac{1}{\G{1-n_1+n_2-n_3+\mu_2+\mu_5-D/2}} \\
\frac{1}{\G{1+n_3+n_4+n_5-\mu_1-\mu_2+D/2}}
\end{multline}
\begin{multline}
\label{psi53}
\{ 1,4,5,6,7\}:\quad
\Psi_3^{(5)}=
|x_{12}|^{D-2(\mu_1+\mu_2)} 
|x_{15}|^{2\mu_2-D} 
|x_{23}|^{-2 \mu_3} 
|x_{24}|^{-2 \mu_4} 
|x_{25}|^{D-2(\mu_2+\mu_5)}
\\
\sum_{\mathbf{n}\in\Z_{\geqslant 0}^5} 
\frac{ 
(\xi^2)^{n_1} 
(\xi^2/\xi^1)^{n_2}
(\xi^2\xi^3/\xi^1)^{n_3} 
(\xi^4)^{n_4} 
(\xi^5/\xi^1)^{n_5}}{\G{1+n_1} \G{1+n_2} \G{1+n_3} \G{1+n_4} \G{1+n_5} }
\\
\frac{1}{\G{1-n_1-n_3-n_4-\mu_3} \G{1-n_2-n_3-n_5-\mu_4}}\\
\frac{1}{\G{1-n_1-n_2-n_3-n_4-n_5+\mu_2-D/2}} 
\frac{1}{\G{1+n_1+n_2+n_3-\mu_2-\mu_5+D/2}} \\
\frac{1}{\G{1+n_3+n_4+n_5-\mu_1-\mu_2+D/2}}
\end{multline}
\begin{multline}
\label{psi54}
\{ 2,3,4,5,8\}:\quad
\Psi_4^{(5)}=
|x_{13}|^{D-2(\mu_1+\mu_3)} 
|x_{14}|^{2(\mu_3+\mu_5)-D} 
|x_{15}|^{-2 \mu_5} 
|x_{23}|^{-2 \mu_2} 
|x_{34}|^{2(\mu_1+\mu_2)-D}
\\
\sum_{\mathbf{n}\in\Z_{\geqslant 0}^5} 
\frac{
(\xi^3)^{n_1} 
(\xi^1)^{n_2} 
(\xi^2)^{n_3} 
(\xi^4/\xi^3)^{n_4} 
(\xi^5/\xi^3)^{n_5}}{\G{1+n_1}\G{1+n_2}\G{1+n_3}\G{1+n_4}\G{1+n_5}}
\\
\frac{1}{\G{1-n_1-n_2-n_3-\mu_2}\G{1-n_3-n_4-n_5-\mu_5}}\\
\frac{1}{\G{1+n_1-n_4-n_5+\mu_1+\mu_2-D/2}} 
\frac{1}{\G{1-n_1-n_2+n_4+\mu_3+\mu_5-D/2}} \\
\frac{1}{\G{1+n_2+n_3+n_5-\mu_1-\mu_3+D/2}}
\end{multline}
\begin{multline}
\label{psi55}
\{ 3,4,5,6,8\}:\quad
\Psi_5^{(5)}=
|x_{14}|^{2(\mu_5-\mu_1)} 
|x_{15}|^{-2\mu_5} 
|x_{23}|^{2(\mu_4+\mu_5)-D} 
|x_{24}|^{2(\mu_1+\mu_3)-D} 
|x_{34}|^{2(\mu_1+\mu_2)-D}
\\
\sum_{\mathbf{n}\in\Z_{\geqslant 0}^5} 
\frac{
(\xi^3)^{n_1} 
(\xi^1)^{n_2} 
(\xi^2/\xi^1)^{n_3} 
(\xi^4/\xi^3)^{n_4} 
(\xi^5/(\xi^1\xi^3))^{n_5}}{\G{1+n_1} \G{1+n_2} \G{1+n_3} \G{1+n_4} \G{1+n_5}}
\\
\frac{1}{\G{1-n_3-n_4-n_5-\mu_5} 
\G{1+n_1-n_4-n_5+\mu_1+\mu_2-D/2}} \\
\frac{1}{\G{1+n_2-n_3-n_5+\mu_1+\mu_3-D/2}} 
\frac{1}{\G{1-n_1-n_2+n_3+n_4+n_5+\mu_5-\mu_1}} \\
\frac{1}{\G{1-n_1-n_2+n_5+\mu_4+\mu_5-D/2}}
\end{multline}
\begin{multline}
\label{psi56}
\{ 4,5,6,7,8\}:\quad  
\Psi_6^{(5)}=
|x_{15}|^{-2\mu_1} 
|x_{23}|^{2(\mu_4+\mu_5)-D} 
|x_{24}|^{2(\mu_3+\mu_5)-D} 
|x_{25}|^{2(\mu_1-\mu_5)} 
|x_{34}|^{2(\mu_1+\mu_2)-D}
\\
\sum_{\mathbf{n}\in\Z_{\geqslant 0}^5} 
\frac{(\xi^2\xi^3/\xi^1)^{n_1} 
(\xi^2)^{n_2} 
(\xi^2/\xi^1)^{n_3} 
(\xi^1\xi^4/(\xi^2\xi^3))^{n_4} 
(\xi^5/(\xi^2\xi^3))^{n_5}}{\G{1+n_1} \G{1+n_2} \G{1+n_3} \G{1+n_4} \G{1+n_5}}
\\
\frac{1}{\G{1-n_1-n_2-n_3-\mu_1}\G{1+n_1+n_2+n_3-n_4-n_5+\mu_1-\mu_5}}\\
\frac{1}{\G{1+n_1-n_4-n_5+\mu_1+\mu_2-D/2}} 
\frac{1}{\G{1-n_1-n_3+n_4+\mu_3+\mu_5-D/2}}\\
\frac{1}{\G{1-n_1-n_2+n_5+\mu_4+\mu_5-D/2}}
\end{multline}
\begin{multline}
\label{psi57}
\{ 2,4,5,8,9\}:\quad  
\Psi_7^{(5)}=
|x_{13}|^{D-2(\mu_1+\mu_3)} 
|x_{15}|^{2\mu_3-D} 
|x_{23}|^{-2 \mu_2} 
|x_{34}|^{-2 \mu_4} 
|x_{35}|^{D-2(\mu_3+\mu_5)}
\\
\sum_{\mathbf{n}\in\Z_{\geqslant 0}^5} 
\frac{
(\xi^4)^{n_1} 
(\xi^4/\xi^3)^{n_2} 
(\xi^1\xi^4/\xi^3)^{n_3} 
(\xi^2)^{n_4} 
(\xi^5/\xi^3)^{n_5}}{\G{1+n_1} \G{1+n_2} \G{1+n_3} \G{1+n_4} \G{1+n_5}}
\\
\frac{1}{\G{1-n_1-n_3-n_4-\mu_2} \G{1-n_2-n_3-n_5-\mu_4}}\\
\frac{1}{\G{1-n_1-n_2-n_3-n_4-n_5+\mu_3-D/2}}
\frac{1}{\G{1+n_1+n_2+n_3-\mu_3-\mu_5+D/2}} \\
\frac{1}{\G{1+n_3+n_4+n_5-\mu_1-\mu_3+D/2}}
\end{multline}
\begin{multline}
\label{psi58}
\{ 4,5,7,8,9\}:\quad  
\Psi_8^{(5)}=
|x_{15}|^{-2\mu_1} 
|x_{23}|^{2(\mu_4+\mu_5)-D} 
|x_{25}|^{2(\mu_1+\mu_3)-D} 
|x_{34}|^{-2 \mu_4} 
|x_{35}|^{D-\mu_3-\mu_5}
\\
\sum_{\mathbf{n}\in\Z_{\geqslant 0}^5} 
\frac{(\xi^4)^{n_1} 
(\xi^2)^{n_2} 
(\xi^4/\xi^3)^{n_3} 
(\xi^1\xi^4/(\xi^2\xi^3))^{n_4} 
(\xi^5/(\xi^2\xi^3))^{n_5}}{\G{1+n_1} \G{1+n_2} \G{1+n_3} \G{1+n_4} \G{1+n_5}}
\\
\frac{1}{\G{1-n_1-n_2-n_3-\mu_1} \G{1-n_3-n_4-n_5-\mu_4}}\\
\frac{1}{\G{1+n_2-n_4-n_5+\mu_1+\mu_3-D/2}}
\frac{1}{\G{1+n_1+n_3+n_4-\mu_3-\mu_5+D/2}}\\
\frac{1}{\G{1-n_1-n_2+n_5+\mu_4+\mu_5-D/2}}
\end{multline}
\begin{multline}
\label{psi59}
\{ 3,4,6,8,10\}:\quad  
\Psi_9^{(5)}=
|x_{14}|^{D-2(\mu_1+\mu_4)} 
|x_{15}|^{2\mu_4-D} 
|x_{24}|^{-2 \mu_2} 
|x_{34}|^{-2 \mu_3} 
|x_{45}|^{D-2(\mu_4+\mu_5)}
\\
\sum_{\mathbf{n}\in\Z_{\geqslant 0}^5} 
\frac{(\xi^5/\xi^1)^{n_1} 
(\xi^5/\xi^3)^{n_2} 
(\xi^5/(\xi^1\xi^3))^{n_3} 
(\xi^2/\xi^1)^{n_4} 
(\xi^4/\xi^3)^{n_5}}{\G{1+n_1} \G{1+n_2} \G{1+n_3} \G{1+n_4} \G{1+n_5}}
\\
\frac{1}{\G{1-n_1-n_3-n_4-\mu_2} \G{1-n_2-n_3-n_5-\mu_3}}\\
\frac{1}{\G{1+n_1+n_2+n_3-\mu_4-\mu_5+D/2}}
\frac{1}{\G{1-n_1-n_2-n_3-n_4-n_5+\mu_4-D/2}} \\
\frac{1}{\G{1+n_3+n_4+n_5-\mu_1-\mu_4+D/2}}
\end{multline}
\begin{multline}
\label{psi510}
\{ 4,6,7,8,10\}:\quad  
\Psi_{10}^{(5)}=
|x_{15}|^{-2 \mu_1} 
|x_{24}|^{2(\mu_3+\mu_5)-D} 
|x_{25}|^{2(\mu_1+\mu_4)-D} 
|x_{34}|^{-2 \mu_3} 
|x_{45}|^{D-2(\mu_4+\mu_5)}
\\
\sum_{\mathbf{n}\in\Z_{\geqslant 0}^5} 
\frac{(\xi^5/\xi^1)^{n_1} 
(\xi^5/\xi^3)^{n_2} 
(\xi^2/\xi^1)^{n_3} 
(\xi^5/(\xi^2\xi^3))^{n_4} 
(\xi^1\xi^4/(\xi^2\xi^3))^{n_5}}{%
\G{1+n_1} \G{1+n_2} \G{1+n_3} \G{1+n_4} \G{1+n_5}} 
\\
\frac{1}{\G{1-n_1-n_2-n_3-\mu_1} \G{1-n_2-n_4-n_5-\mu_3}}\\
\frac{1}{\G{1+n_1+n_2+n_4-\mu_4-\mu_5+D/2}}
\frac{1}{\G{1+n_3-n_4-n_5+\mu_1+\mu_4-D/2}}\\
\frac{1}{\G{1-n_1-n_3+n_5+\mu_3+\mu_5-D/2}}
\end{multline}
\begin{multline}
\label{psi511}
\{ 4,7,8,9,10\}:\quad
\Psi_{11}^{(5)}=
|x_{15}|^{-2 \mu_1} 
|x_{25}|^{-2 \mu_2} 
|x_{34}|^{2\mu_5-D} 
|x_{35}|^{D-2(\mu_3+\mu_5)} 
|x_{45}|^{D-2(\mu_4+\mu_5)}
\\
\sum_{\mathbf{n}\in\Z_{\geqslant 0}^5} 
\frac{(\xi^4\xi^5/(\xi^2\xi^3))^{n_1} 
(\xi^5/\xi^3)^{n_2} 
(\xi^4/\xi^3)^{n_3} 
(\xi^5/(\xi^2\xi^3))^{n_4} 
(\xi^1\xi^4/(\xi^2\xi^3))^{n_5}}{\G{1+n_1} \G{1+n_2} \G{1+n_3} \G{1+n_4} \G{1+n_5}} 
\\
\frac{1}{\G{1-n_1-n_2-n_3-\mu_1}\G{1-n_1-n_4-n_5-\mu_2}}\\
\frac{1}{\G{1+n_1+n_2+n_4-\mu_4-\mu_5+D/2}}
\frac{1}{\G{1+n_1+n_3+n_5-\mu_3-\mu_5+D/2}}\\
\frac{1}{\G{1-n_1-n_2-n_3-n_4-n_5+\mu_5-D/2}}
\end{multline}
The coefficients of the $\Gamma$-series in the permutation-symmetric solution
are as in \eq{cigma}. For example, the coefficient of $\Psi^{(5)}_1$ is 
given by 
\begin{equation}
	C_1^{-1} = \sin \pi\mu_4 \sin\pi\mu_5 \sin\pi (D/2-\mu_1)
\sin\pi (\mu_1+\mu_3-D/2)\sin\pi(\mu_1+\mu_2-D/2),
\end{equation}
as may be seen from the denominator of the $\Gamma$-series \eq{psi51}.
\end{appendices}
\clearpage
\printbibliography
\end{document}